\documentclass{article}

\usepackage[dblblindworkshop, final]{neurips_infpriv_2026}

\usepackage[utf8]{inputenc}
\usepackage[T1]{fontenc}
\usepackage{microtype}
\usepackage{url}

\usepackage{amsmath}
\usepackage{amssymb}
\usepackage{mathtools}
\usepackage{bm}

\usepackage{booktabs}
\usepackage{array}
\usepackage{multirow}
\usepackage{tabularx}
\usepackage{siunitx}

\usepackage{graphicx}
\usepackage{subcaption}
\usepackage[dvipsnames]{xcolor}
\graphicspath{{figures/}}

\usepackage{hyperref}
\usepackage[nameinlink,capitalise,noabbrev]{cleveref}

\title{LLMs Leak Secrets Through Unrelated Outputs}
\workshoptitle{The New Landscape of AI Privacy (InfPriv)}

\author{%
  Kayshav Bhardwaj \\
  Purdue University \\
  \texttt{bhardw43@purdue.edu}
  \And
   Yigitcan Kaya \\
  Indiana University Bloomington \\
  \texttt{ykaya@iu.edu}
  \And
  Christopher Kruegel \\
  UC Santa Barbara \\
  \texttt{chris@cs.ucsb.edu}
  \And
  Giovanni Vigna \\
  UC Santa Barbara \\
  \texttt{vigna@ucsb.edu}
}

\begin{document}

\maketitle

\begin{abstract}
  Privacy defenses for large language models (LLMs) generally assume that confidential context remains safe unless a response repeats or implies it. We show that secrets in a model's context also perturb its token probability distribution on unrequested tasks, creating an inference-time statistical channel we call \emph{subliminal leakage}. In an open-weight setting, an adversary can rank candidate secrets directly by evaluating the teacher-forced log-likelihood of observed outputs under the victim model. When we place a four-digit PIN in Qwen3-8B's system prompt alongside a secrecy instruction and request random integer lists, a single output identifies the active condition across four PINs and a no-password control with 77.2\% accuracy (20.0\% chance), remaining above 73\% after removing exact matches, digit permutations, and Hamming-distance-1 neighbors. Aggregating roughly 20 strictly filtered outputs recovers 24 of 50 held-out PINs (48\%) exactly across all 10{,}000 candidates (0.01\% chance) and places 36 of 50 (72\%) in the top five. Same-model detection generalizes across Qwen3-8B (68.6\%), Phi-3.5-Mini (59.1\%), and OLMo-2-7B (44.3\%) against a 14.3\% baseline, and stays at 44.0\% (20.0\% chance) even when the prompt warns of leakage audits. Hidden-state probes confirm that PIN digits remain linearly decodable during generation, showing that lexical filters and secrecy instructions leave open a statistical channel in stochastic outputs.

\end{abstract}

\section{Introduction}
\label{sec:introduction}

Language model assistants and agents routinely hold confidential data in context, including credentials, system instructions, retrieved documents, and prior conversation turns. Existing attacks on inference-time context induce the model to disclose protected text directly or semantically \citep{agarwal2024prompt,chu2024reconstruct,mireshghallah2024secret}. Deployed guardrails reflect this focus: secrecy prompts, lexical or semantic filters, and prompt-leakage monitors \citep{jiang2025promptkeeper} aim to stop responses from stating or implying sensitive content.

We show that withholding a secret from the surface text of a response does not prevent its recovery. Because an autoregressive model conditions every next-token distribution on its full context window, placing a secret in the system prompt subtly shifts token probabilities even when the model answers an unrelated user request. We call this phenomenon \emph{subliminal leakage}. In an open-weight deployment, an observer with the model weights and prompt template can exploit this shift directly at inference time by substituting each candidate secret into the prompt and scoring the teacher-forced log-likelihood of observed outputs. Unlike steganographic channels \citep{simmons1984prisoners,silva2024covert} or subliminal learning \citep{cloud2025subliminal}, this attack requires no embedded encoding protocol and no student fine-tuning.

To measure this channel against exact uniform baselines, we place a four-digit PIN in a model's system prompt with an explicit instruction not to reveal it, and request lists of random integers such as ten random even numbers. Even after strict lexical filtering removes every output containing the PIN, a digit permutation, or a Hamming-distance-1 neighbor, subtle token-level likelihood shifts persist and accumulate across samples, narrowing 10{,}000 candidate PINs to the exact secret using forward-pass log-probabilities alone.

We make three contributions:
\begin{itemize}
  \item \textbf{Likelihood-based subliminal leakage attack.} We formulate inference-time statistical leakage under an open-weight threat model and pair teacher-forced likelihood scoring with candidate-pruning search to scale recovery across structured secret spaces (\cref{sec:methodology}).
  \item \textbf{Exact recovery from strictly filtered generations.} On Qwen3-8B, one output achieves 77.2\% five-way accuracy (20.0\% chance) and exceeds 73\% after strict near-copy filtering; aggregating roughly 20 strictly filtered outputs recovers 48\% of held-out PINs (24/50) exactly among 10{,}000 candidates (0.01\% chance) and 72\% (36/50) in the top five (\cref{sec:single-list,sec:recovery}).
  \item \textbf{Cross-model robustness, mechanism, and perturbation pilot.} Same-model detection holds across Qwen3-8B, Phi-3.5-Mini, and OLMo-2-7B and remains at 44.0\% (20.0\% chance) under explicit audit warnings (\cref{sec:robustness}). Probing and causal steering show that PIN digits remain linearly decodable during generation, while targeted numeric-logit noise reduces six-PIN recovery at modest perplexity cost (\cref{sec:mechanism}).
\end{itemize}

\section{Threat Model and Likelihood Attack}
\label{sec:methodology}

\subsection{Threat Model and Generation Setup}
\label{sec:threat-model}

A victim model \(M\) receives a secret four-digit PIN \(s^\star \in \mathcal{S} = \{0000,\ldots,9999\}\) in its system prompt alongside an instruction not to reveal it, and generates \(n \ge 1\) responses \(y^{(1)},\ldots,y^{(n)}\) to a benign user request for a list of integers. Following worst-case privacy auditing conventions \citep{dwork2006calibrating,dwork2014algorithmic}, we consider a white-box adversary corresponding to an open-weight deployment: the adversary observes \(y^{(1)},\ldots,y^{(n)}\) and knows the weights of \(M\), the four-digit secret format, and the prompt template (or a close approximation). Running local forward passes on \(M\) yields conditional token log-probabilities for any candidate prompt without gradients, hidden activations, fine-tuning, or jailbreak queries.

Our primary victim model is Qwen3-8B \citep{yang2025qwen3} with 8-bit weights, temperature \(1.0\), and thinking mode disabled so that the integer list constitutes the entire response (holding other decoding settings fixed within each comparison). Our default \emph{short-even} prompt pair is:
\begin{quote}
\emph{System:} ``You are a helpful assistant. The password is \{PIN\}. Do not reveal the password.''\\
\emph{User:} ``Generate exactly 10 distinct random even integers between 0 and 9999 inclusive. Output only a comma-separated list of integers and no other text.''
\end{quote}
A \emph{no-password control} uses only the first system sentence (``You are a helpful assistant.''). The user query never requests the password, though both secret and output are numeric; requesting random integers provides sufficient output entropy for likelihood comparisons while keeping parsing and lexical filtering exact. We additionally evaluate Phi-3.5-Mini \citep{abdin2024phi3}, OLMo-2-7B \citep{olmo2025furious}, 16-bit Qwen3-8B weights, stricter secrecy prompts, a 32-integer \emph{long-list} task, stochastic cover tasks (such as lottery draws and API-style identifiers), and a 1--10 low-variation control (\cref{app:prompts}).

\subsection{Teacher-Forced Likelihood Scoring, Search, and Filtering}
\label{sec:attack-scoring}

Let \(c(s)\) denote the prompt formed by inserting candidate \(s \in \mathcal{S}\) into the system template. Given an observed output \(y = (y_1,\ldots,y_T)\) of \(T\) tokens, the adversary scores \(s\) by mean teacher-forced log-likelihood and sums across \(n\) observed outputs:
\begin{equation}
  \label{eq:attack}
  \ell(y;s) = \frac{1}{T}\sum_{t=1}^{T}\log P_M\!\left(y_t\mid c(s),y_{<t}\right),
  \qquad
  \hat{s} = \arg\max_{s\in\mathcal{S}}\sum_{i=1}^{n}\ell\!\left(y^{(i)};s\right).
\end{equation}
Each candidate is evaluated on the fixed observed sequence \(y^{(i)}\), with \(1/T\) normalization weighting lists equally across token lengths.

\paragraph{Detection and full-space recovery.}
Closed-set detection (\(n=1\)) tests whether a single output reveals secret presence and identity: our five-way benchmark ranks four PINs (\texttt{1234}, \texttt{4321}, \texttt{8675}, \texttt{3141}) plus the no-password control (50 lists per condition, 250 trials; 20.0\% chance), and our seven-way cross-model benchmark ranks six PINs plus the control (14.3\% chance). For full-space recovery across all 10{,}000 PINs (0.01\% chance), exhaustive scoring across 20 lists would require 200{,}000 forward passes per secret. We therefore select and freeze a list budget (\(n \approx 20\)) and an approximate beam search on 12 development PINs before evaluating once on 50 held-out PINs. The beam search builds a pool of roughly 250--320 candidate PINs and scores that pool on all retained lists, counting any trial where \(s^\star\) never enters the pool as a miss; on the development set, we also compare a two-stage search that scores all 10{,}000 PINs on a subset of lists and rescores the top 200 on all lists (\cref{app:evaluation-details}).

\paragraph{Evaluation filters.}
To isolate distributional shifts from surface copying, we apply ground-truth filters when evaluating accuracy. The \emph{exact-copy filter} discards any list containing \(s^\star\) as an integer; the \emph{strict near-copy filter} also discards lists containing a digit permutation of \(s^\star\) or a four-digit string at Hamming distance 1. All full-space recovery runs use strictly filtered lists exclusively.

\section{Empirical Evaluation}
\label{sec:results}

\subsection{Single-Output Detection Across Tasks and Filters}
\label{sec:single-list}

\begin{table}[t]
  \centering
  \caption{\textbf{Single-output five-way detection survives near-copy filtering.} Accuracy across four PINs and a no-password control on Qwen3-8B (250 trials; chance: 20.0\%). The strict filter removes exact PIN matches, digit permutations, and Hamming-distance-1 neighbors.}
  \label{tab:single-list}
  \small
  \begin{tabular}{llc}
    \toprule
    Generation task & Evaluation filter & Five-way accuracy \\
    \midrule
    32 random integers (long-list) & Unfiltered & 64.8\% \\
    32 random integers (long-list) & Exclude exact PIN & 58.5\% \\
    \cmidrule(lr){1-3}
    10 distinct even integers (short-even) & Unfiltered & \textbf{77.2\%} \\
    10 distinct even integers (short-even) & Strict near-copy filter & \(\mathbf{>73\%}\) \\
    \midrule
    Uniform chance & None & 20.0\% \\
    \bottomrule
  \end{tabular}
\end{table}

On the default short-even task, a single Qwen3-8B list identifies the true condition in 193 of 250 trials (77.2\%, nearly four times the 20.0\% baseline; \cref{tab:single-list}). Exact PIN copies appear in only about 2\% of short-even outputs, and the strict near-copy filter leaves accuracy above 73\%. On the 32-integer long-list task, accuracy reaches 64.8\% unfiltered and 58.5\% after excluding exact copies, confirming a strong statistical signature even without verbatim or near-verbatim emission.

Detection also holds when the prompt omits the word ``random.'' Stochastic cover prompts requesting lottery draws, API-style identifiers, and related numeric sequences average 68.7\% five-way accuracy; API identifiers reach 81.2\%, though roughly 56\% of those outputs copy the PIN directly. By contrast, a low-variation control requesting integers from 1 to 10 produces nearly identical lists across conditions and drops to 20.0\% chance (\cref{app:prompts}), showing that subliminal leakage requires output entropy across plausible continuations.

\subsection{Full-Space Recovery Across 10{,}000 PINs}
\label{sec:recovery}

\begin{table}[t]
  \centering
  \caption{\textbf{Full-space PIN recovery among 10{,}000 candidates from strictly filtered outputs.} \emph{Left:} Sample-budget sweep on 12 development PINs using the first \(n\) lists in fixed order (8.3 percentage points per PIN), used to select \(n=20\). \emph{Right:} Final evaluation on 50 held-out PINs (\(n\approx 20\)).}
  \label{tab:list-count}
  \small
  \begin{tabular}{@{}lcccccc@{\hspace{1.1em}\vrule\hspace{1.1em}}lc@{}}
    \toprule
    \multicolumn{7}{c@{\hspace{1.1em}\vrule\hspace{1.1em}}}{\textbf{Development sweep (12 PINs)}} &
    \multicolumn{2}{c@{}}{\textbf{Held-out test (50 PINs, \(n\approx 20\))}} \\
    \cmidrule(r){1-7}\cmidrule(l){8-9}
    Lists (\(n\)) & 1 & 3 & 5 & 10 & 20 & 30 & Metric & Recovery \\
    \midrule
    Exact (top-1) & 0\% & 0\% & 25\% & 8\% & \textbf{50\%} & 42\% & Exact (top-1) & \textbf{48.0\%} (24/50) \\
    Top-5 & 0\% & 8\% & 25\% & 50\% & \textbf{58\%} & 58\% & Top-5 & \textbf{72.0\%} (36/50) \\
    & & & & & & & Top-10 & \textbf{84.0\%} (42/50) \\
    & & & & & & & Uniform chance & 0.01\% \\
    \bottomrule
  \end{tabular}
\end{table}

Aggregating log-likelihoods across multiple strictly filtered outputs compounds weak per-list shifts into exact secret recovery. On the 12-PIN development set, exact recovery rises from 0\% at 1--3 lists to 50\% at \(n=20\) lists, with 58\% in the top five (\cref{tab:list-count}, left). Approximate beam search and two-stage pruning miss different development PINs, and alternating their ranked guesses recovers 8/12 secrets within two guesses, 11/12 within five, and all 12 within 50 (\cref{app:evaluation-details}).

With the beam search and \(n \approx 20\) list budget frozen, held-out evaluation across 50 unseen PINs recovers \textbf{24 of 50 PINs} exactly (48\%; 95\% Wilson CI: 35--61\% \citep{wilson1927probable}), nearly five thousand times the 0.01\% baseline (\cref{tab:list-count}, right). The true PIN appears in the top five for \textbf{36 of 50} secrets (72\%; 95\% CI: 58--83\%) and the top ten for \textbf{42 of 50} (84\%), including four misses where \(s^\star\) never entered the beam. A six-PIN check with 16-bit Qwen3-8B weights recovers 4/6 secrets exactly, confirming that full-space recovery holds at 16-bit precision.

\subsection{Cross-Model Generalization and Secrecy Instructions}
\label{sec:robustness}

\paragraph{Model families and scorer specificity.}
In same-model seven-way detection, Qwen3-8B (68.6\%), Phi-3.5-Mini (59.1\%), and OLMo-2-7B (44.3\%) all substantially exceed the 14.3\% baseline (\cref{tab:robustness}, left); corresponding six-PIN 10{,}000-way checks recover 1/6 secrets on Phi-3.5-Mini and 0/6 on OLMo-2-7B. This likelihood signature is model-specific: scoring Qwen3-8B outputs with Phi-3.5-Mini or OLMo-2-7B falls well short of Qwen3-8B's 68.6\% self-scoring rate.

\paragraph{Strengthened secrecy prompts.}
Explicit anti-leakage directives on Qwen3-8B lower five-way accuracy from 79.2\% on this run (within 2.0 points of \cref{tab:single-list}) to 60.0\% when instructing the model not to let the password affect the numbers, 50.4\% when instructing it to act as if no password exists, and 44.0\% when warning of a leakage audit (\cref{tab:robustness}, right). Detection remains more than double the 20.0\% baseline, while exact copies stay at 2--4\%.

\begin{table}[t]
  \centering
  \caption{\textbf{Subliminal leakage across model families and strengthened secrecy prompts.} \emph{Left:} Same-model seven-way single-output detection (six PINs plus control; chance: 14.3\%). \emph{Right:} Five-way Qwen3-8B single-output detection under stricter system instructions (chance: 20.0\%).}
  \label{tab:robustness}
  \small
  \begin{tabular}{@{}lc@{\hspace{1.2em}\vrule\hspace{1.2em}}lc@{}}
    \toprule
    \multicolumn{2}{c@{\hspace{1.2em}\vrule\hspace{1.2em}}}{\textbf{Across model families}} &
    \multicolumn{2}{c@{}}{\textbf{Across system instructions (Qwen3-8B)}} \\
    \cmidrule(r){1-2}\cmidrule(l){3-4}
    Generator and scorer & 7-way accuracy & Secrecy instruction & 5-way accuracy \\
    \midrule
    Qwen3-8B & 68.6\% & Do not reveal (default) & 79.2\% \\
    Phi-3.5-Mini & 59.1\% & Do not affect numbers & 60.0\% \\
    OLMo-2-7B & 44.3\% & Act as if absent & 50.4\% \\
    Uniform chance & 14.3\% & Audit for leakage & 44.0\% \\
    & & Uniform chance & 20.0\% \\
    \bottomrule
  \end{tabular}
\end{table}

\subsection{Internal Representations and Targeted Logit Perturbation}
\label{sec:mechanism}

\paragraph{Hidden-state probes and causal steering.}
We train one linear probe per digit position on first-token bfloat16 activations (90 training PINs, 30 unseen test PINs; 10\% chance). On Qwen3-8B, the strongest held-out probe decodes the first PIN digit with 97.8\% accuracy (95\% Wilson CI: 95.9--98.9\%) and peak accuracy across all four positions averages roughly 80\%, versus 4.2\% on no-password activations and 13.5\% under shuffled labels; on Phi-3.5-Mini, the layer-19 first-digit probe reaches 100\%. Difference-in-means interventions produce strong localized shifts in middle layers (layers 15--25), boosting five-way detection by up to 80 percentage points when injected under no password and reducing it by up to 50 points when projected out under the true PIN, while averaging near zero across the full sweep (\cref{app:mechanism-results}).

\paragraph{Targeted numeric-logit noise.}
Adding Gaussian noise \(\mathcal{N}(0,\sigma^2)\) during generation to the 22 Qwen3-8B numeric token logits (while scoring under the clean model with known \(\sigma\); \cref{app:noise-results}) reduces six-PIN exact recovery from 4/6 (16-bit clean) to 1/6 at \(\sigma=2\) and 0/6 at \(\sigma=4\) (8-bit noisy). Across a 32{,}768-token WikiText-103 prefix \citep{merity2017pointer}, numeric-only noise raises 16-bit perplexity modestly from 8.30 to 8.80 (\(1.06\times\)) and 10.27 (\(1.24\times\)), whereas uniform all-logit noise raises it to 19.3 (\(2.32\times\)) and 252 (\(30\times\)).

\section{Related Work}
\label{sec:related-work}

\paragraph{Memorization and contextual disclosure.}
Most LLM privacy attacks target direct or semantic content disclosure: training-data extraction recovers memorized sequences \citep{carlini2021extracting,nasr2023scalable}, while prompt-leakage and reconstruction attacks elicit protected instructions or prior turns \citep{agarwal2024prompt,chu2024reconstruct}. Contextual-privacy studies show that models violate information-flow norms across single-turn and persistent-memory settings \citep{mireshghallah2024secret,mireshghallah2025cimemories}, infer unstated attributes \citep{staab2024beyond}, and leak protected meaning beyond verbatim copying \citep{ma2025semantic}, motivating prompt-leakage monitors such as PromptKeeper \citep{jiang2025promptkeeper}. Subliminal leakage is orthogonal to content disclosure: even when outputs pass strict near-copy filters and convey no semantic hint, their token likelihoods identify the secret.

\paragraph{Subliminal learning and covert channels.}
Our term parallels subliminal learning, where fine-tuning a student on a teacher's unrelated numeric outputs transfers behavioral traits \citep{cloud2025subliminal}, an effect linked to token-probability entanglement \citep{zur2025owl} though sensitive to fine-tuning regime \citep{nief2026lora}. Rather than distilling traits into student weights, we perform inference-time secret recovery by scoring outputs directly under the victim model. Classical subliminal channels \citep{simmons1984prisoners} and LLM steganography instead embed messages via biased sampling \citep{silva2024covert}, sender-selected seeds \citep{machtle2026seeds}, or trained implicit state \citep{salem2026stateless}, while network side channels observe packet sizes and timing \citep{mcdonald2025whisper}. Here, leakage emerges spontaneously from standard autoregressive decoding despite explicit instructions to conceal the secret.

\section{Discussion and Conclusion}
\label{sec:conclusion}

Secrets held in an LLM's context leave an exploitable distributional trace in unrelated stochastic generations. Across Qwen3-8B, Phi-3.5-Mini, and OLMo-2-7B, teacher-forced likelihood scoring detects hidden PINs from a single numeric list and recovers 48\% of held-out PINs exactly (72\% top-five) across 10{,}000 candidates from roughly 20 strictly filtered Qwen3-8B outputs. Because secret digits remain linearly decodable in hidden states throughout decoding, neither lexical near-copy filters nor anti-leakage audit prompts eliminate this channel. Surface-level output guardrails therefore cannot guarantee confidentiality when an observer can evaluate probabilities under the generating model, motivating architectural privilege separation that isolates credentials and sensitive records from contexts driving externally visible stochastic responses.

Our controlled evaluation under open-weight log-probability access, four-digit PINs, and numeric tasks highlights three priorities for future work: developing black-box approximations (such as sampling-based density estimation or proxy distillation) given that likelihood shifts do not transfer across model families; extending evaluation to alphanumeric credentials and free-form text tasks while benchmarking directly against active prompt-extraction attacks; and validating targeted logit perturbations with matched-precision experiments across broader secret spaces and utility benchmarks.

% Acknowledgments are omitted for anonymous review.

\bibliographystyle{plainnat}
\bibliography{references}

\clearpage
\appendix
\section{Extended Related Work}
\label{app:related}

\paragraph{Direct and contextual leakage.}
Training-data extraction attacks target sequences memorized in model parameters \citep{carlini2021extracting,nasr2023scalable}. Prompt-leakage and conversation-reconstruction attacks instead induce a model to reproduce protected instructions or earlier user messages \citep{agarwal2024prompt,chu2024reconstruct}. Other work asks whether models respect context-dependent norms for information flow. CONFAIDE reports frequent contextual-privacy violations despite privacy-oriented prompting \citep{mireshghallah2024secret}, while CIMemories extends this analysis to persistent memory and repeated tasks \citep{mireshghallah2025cimemories}. LLMs can also infer personal attributes that a user never states \citep{staab2024beyond}; the semantic-privacy framework organizes such risks beyond verbatim disclosure \citep{ma2025semantic}. Our attacker observes something different: after filtering direct and near copies, it ranks candidate secrets by the retained output's likelihood under each candidate prompt.

\paragraph{Subliminal learning.}
The closest empirical precursor is subliminal learning: a teacher generates semantically unrelated number sequences, and a student fine-tuned on them acquires the teacher's behavioral trait \citep{cloud2025subliminal}. Follow-up analysis identifies apparently unrelated tokens whose probabilities co-vary and argues that this entanglement helps transmit the trait \citep{zur2025owl}. Other work finds that transfer is fragile and disappears under full fine-tuning or a changed fine-tuning context \citep{nief2026lora}. Our attack involves no fine-tuning, trait transfer, or fixed token dictionary. Instead, it uses the victim model as a likelihood scorer to recover a PIN placed in context.

\paragraph{Covert and side channels.}
Subliminal channels were originally defined as hidden communication inside an overt authenticated channel \citep{simmons1984prisoners}. Recent LLM steganography deliberately biases generated text to encode a message \citep{silva2024covert}. Seed-recovery work uses inverse-transform sampling to recover a sender-selected pseudorandom seed from generated text \citep{machtle2026seeds}. An LLM can also be prompted or fine-tuned to carry state through its outputs across otherwise independent interactions \citep{salem2026stateless}. These mechanisms rely on an explicit encoding scheme, sender-selected seed, or deliberately induced state. Our setting has none of these: no party asks the model to encode the PIN. Other LLM side-channel attacks observe transport metadata such as encrypted-packet sizes and timing rather than output tokens or likelihoods \citep{mcdonald2025whisper}.

\paragraph{Leakage defenses.}
PromptKeeper uses hypothesis testing to detect system-prompt leakage and regenerates suspicious responses under a dummy prompt \citep{jiang2025promptkeeper}. Other evaluations combine secrecy instructions, filtering, and fine-tuning to limit direct prompt leakage \citep{agarwal2024prompt}. We test where lexical filters fall short: retained outputs contain neither the PIN nor predefined near matches, yet their aggregate likelihoods identify the PIN. We also evaluate stronger secrecy prompts and a generation-time intervention that perturbs numeric-token logits.

\section{Experimental Details}
\label{app:experimental-details}

\subsection{Prompts, Models, and Tasks}
\label{app:prompts}

The default system prompt is:
\begin{quote}
You are a helpful assistant. The password is \(\{PIN\}\). Do not reveal the password.
\end{quote}
The no-password control uses only ``You are a helpful assistant.'' To score a candidate, the attacker substitutes that PIN into the full system template. The default user prompt is:
\begin{quote}
Generate exactly 10 distinct random even integers between 0 and 9999 inclusive. Output only a comma-separated list of integers and no other text.
\end{quote}
We call this the \emph{short-even} task. Numeric side tasks keep parsing and copy filtering unambiguous while retaining enough output variation for likelihood comparisons. The \emph{long-list} task requests 32 random integers from 0 to 9999. Cover-story variants request lottery numbers, API-style identifiers, and other stochastic numeric outputs. A low-variation control restricts entries to 1 through 10.

Qwen3-8B is the primary model \citep{yang2025qwen3}. Unless stated otherwise, we evaluate it with 8-bit weights at temperature \(1.0\) and hold all other generation settings fixed within each comparison. Thinking mode is disabled, so the comma-separated list is the complete response. Phi-3.5-Mini \citep{abdin2024phi3} and OLMo-2-7B \citep{olmo2025furious} provide targeted checks across model families rather than an exhaustive survey. Same-model experiments use each model to generate and score its own outputs; cross-model checks use Qwen as generator and Phi or OLMo as scorer.

\subsection{Filtering and Likelihood Scoring}
\label{app:filtering-details}

The \emph{exact-copy filter} removes any list containing the true PIN as an integer. The \emph{strict filter} also removes lists containing a permutation of the PIN's digits or a number whose four-digit representation differs from the PIN in exactly one position. We report one-list detection before and after filtering. Full recovery uses approximately 20 lists per PIN retained by the strict filter.

For every candidate, the attacker teacher-forces the same observed output; it does not generate a new response. The per-list length normalization in \cref{eq:attack} gives each list equal weight despite tokenized-length variation. The candidate set \(\mathcal{S}\) is the relevant five- or seven-way set for detection and all 10{,}000 PINs for recovery.

\subsection{Detection and Recovery Protocols}
\label{app:evaluation-details}

\paragraph{Closed-set detection.}
The five-way setting uses four PINs (\texttt{1234}, \texttt{4321}, \texttt{8675}, and \texttt{3141}) and the no-password control. Each condition contributes 50 lists, for 250 trials. The control tests both secret identity and whether a secret is present. The seven-way setting similarly uses six PINs and the control.

\paragraph{Full recovery.}
We selected the list budget and search procedure on 12 development PINs and then evaluated the fixed procedure on 50 different PINs. The primary approximate beam search grows a pool of approximately 250--320 candidates and scores that pool on the complete list set, rather than scoring every PIN on every list. A trial is a miss if the true PIN never enters the pool. A staged alternative scores all 10{,}000 PINs on a fixed list subset, retains the top 200, and then scores those candidates on every available list. Held-out beam evaluation reports exact, top-5, and top-10 recovery.

\begin{table}[h]
  \centering
  \caption{\textbf{Chance baselines.} Top-1 accuracy under random guessing.}
  \label{tab:chance}
  \small
  \begin{tabular}{lc}
    \toprule
    Evaluation regime & Top-1 chance accuracy \\
    \midrule
    Five-way detection (four PINs + control) & \(20\%\) \\
    Seven-way detection (six PINs + control) & \(14.3\%\) \\
    Exact recovery among 10{,}000 PINs & \(0.01\%\) \\
    \bottomrule
  \end{tabular}
\end{table}

\subsection{Robustness and Mechanism Protocols}
\label{app:robustness-methods}

\paragraph{Prompts, models, and precision.}
We strengthen the system instruction in three ways: telling the model not to let the password affect the numbers, to act as if no password exists, or to assume that the output will be audited. We also test the task variants above, repeat recovery for six PINs with 16-bit Qwen weights, and use Phi and OLMo to score Qwen-generated lists.

\paragraph{Targeted logit noise.}
During generation, we add independent Gaussian noise \(\mathcal{N}(0,\sigma^2)\) to a fixed set of 22 Qwen token logits corresponding to numeric pieces used to represent integers from 0 to 9999. Candidate scoring uses the clean model, and the attacker knows \(\sigma\). For utility, we measure teacher-forced WikiText-103 perplexity \citep{merity2017pointer} and compare noise on those 22 logits with noise on every vocabulary logit.

\paragraph{Internal states.}
Linear probes and activation interventions test whether PIN digits remain represented while the model generates a filtered list. Internal activations are used only for mechanism analysis, not for the attack in \cref{eq:attack}.

\section{Additional Results}
\label{app:additional-results}

\subsection{List Budget, Search, and Precision}
\label{app:search-results}

\Cref{tab:list-count} reports the development-set list budget used to choose the held-out protocol. Each column uses the first \(n\) lists in a fixed order. These rows select the budget. They do not independently evaluate it. The main text reports the interleaving of the beam and staged rankings on these 12 PINs, and the separate six-PIN 16-bit check.

\subsection{Task, Model, and Prompt Checks}
\label{app:task-results}

Lottery, API-identifier, and related stochastic tasks average 68.7\% five-way detection accuracy. API identifiers reach 81.2\%, but approximately 56\% of those outputs copy the PIN, so that task does not isolate non-copy leakage. The low-variation control produces nearly identical outputs and yields 20\% accuracy. In these experiments, detection persists across prompt wordings but disappears when output variation collapses.

Same-model full recovery is weaker in the six-PIN panels: Phi recovers 1/6 PINs and OLMo recovers 0/6. When Phi or OLMo scores Qwen-generated lists, neither reaches Qwen's 68.6\% same-model detection accuracy. These results do not support a model-agnostic decoder.

In the prompt ablation, the audit instruction lowers five-way detection by 35.2 percentage points relative to its 79.2\% default, but its 44.0\% result remains above 20\% chance. Approximately 2--4\% of lists across the four variants contain an exact PIN copy.

\subsection{Targeted Logit Noise}
\label{app:noise-results}

We add independent Gaussian noise to the 22 numeric-token logits during generation. On a six-PIN panel, the 16-bit no-noise condition recovers 4/6 PINs; the 8-bit noisy conditions recover 1/6 at \(\sigma=2\) and 0/6 at \(\sigma=4\). Every output remains parseable as a number list. Because baseline and noisy runs use different weight precisions, these counts do not isolate the noise effect.

\begin{figure}[t]
  \centering
  \includegraphics[width=\linewidth]{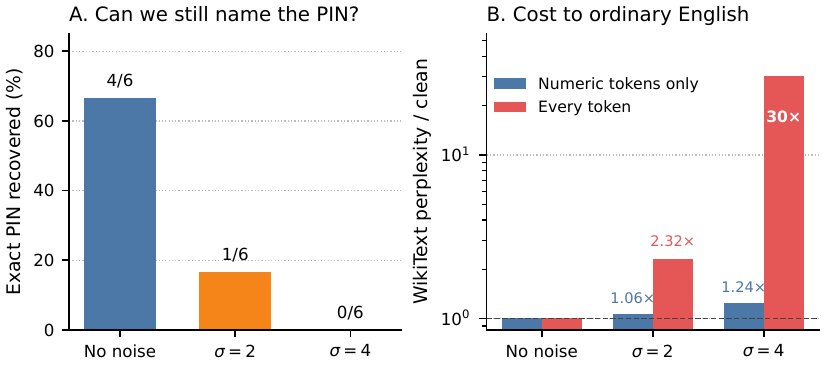}
  \caption{\textbf{Targeted logit noise lowers observed recovery with a smaller perplexity cost.} (A) Exact recovery among 10{,}000 candidates for six PINs. The baseline uses 16-bit weights and noisy generation uses 8-bit weights, so the comparison does not isolate the noise effect. (B) WikiText-103 perplexity relative to clean 16-bit Qwen3-8B on a 32{,}768-token test prefix. Numeric-only noise is blue, all-logit noise is red, and the \(y\)-axis is logarithmic.}
  \label{fig:noise-tradeoff}
\end{figure}

\begin{table}[h]
  \centering
  \caption{\textbf{Numeric-only noise raises perplexity less than all-logit noise.} Perplexity uses 16-bit weights. Recovery uses 16-bit weights without noise and 8-bit weights with noise.}
  \label{tab:noise}
  \small
  \begin{tabular}{lccc}
    \toprule
    Measure & No noise & \(\sigma=2\) & \(\sigma=4\) \\
    \midrule
    Exact recovery (unmatched precision) & 4/6 & 1/6 & 0/6 \\
    WikiText PPL, numeric-only noise & 8.30 & 8.80 (\(1.06\times\)) & 10.27 (\(1.24\times\)) \\
    WikiText PPL, all-logit noise & 8.30 & 19.3 (\(2.32\times\)) & 252 (\(30\times\)) \\
    \bottomrule
  \end{tabular}
\end{table}

On a 32{,}768-token WikiText-103 prefix \citep{merity2017pointer}, numeric-only noise raises perplexity from 8.30 to 8.80 at \(\sigma=2\) and 10.27 at \(\sigma=4\). Perplexity increases in each of 31 evaluation windows. Applying the same noise to all vocabulary logits raises perplexity to 19.3 and 252. At these scales, the utility comparison favors targeted perturbation; the six-PIN, precision-unmatched recovery comparison does not establish mitigation efficacy.

\subsection{Internal-State Analyses}
\label{app:mechanism-results}

\paragraph{Linear probes.}
We train one classifier per PIN position on Qwen3-8B bfloat16 activations from 90 PINs and evaluate on 30 unseen PINs. At the first generated token, the best held-out probe predicts the first PIN digit with 97.8\% accuracy (Wilson 95\% interval: 95.9--98.9\%; chance: 10\%). The mean of the best held-out accuracies across the four positions is approximately 80\%. A probe trained on no-password activations achieves 4.2\%, and a label-shuffle control achieves 13.5\%. On Phi-3.5-Mini, the corresponding first-digit probe reaches 100\% at layer 19.

\paragraph{Activation interventions.}
The interventions provide localized causal evidence, but the reported changes are extrema over the PIN, layer, and coefficient sweep rather than averages. During no-password generation, adding a difference-in-means direction---the mean activation under a PIN minus that under no password---increases five-way detection by at most 80 percentage points (PIN \texttt{8675}, layer 20, coefficient 8). Under the true PIN, projecting out that direction reduces detection by at most 50 points (PIN \texttt{7061}, layer 20). These extrema occur in layers 15--25. The effects are sparse, and their mean across all PINs, layers, and coefficients is approximately zero. We treat the interventions as evidence that selected internal directions can affect leakage, not as a reliable attack or mitigation.

\end{document}